\documentstyle[prb,aps,multicol,epsf]{revtex}

\begin{document}

\draft
\title{Correlation amplitude for $S=\frac{1}{2}$ XXZ spin chain in the
critical region:\\
Numerical renormalization-group study of an open chain
} 
\author{T. Hikihara}
\address{Division of Information and Media Science, Graduate School 
of Science and Technology, \\ Kobe University, Rokkodai, Kobe
657-8501, Japan}
\author{A. Furusaki$^*$}
\address{Department of Physics, Stanford University, Stanford, CA 94305}
\date{13 March 1998}
\maketitle
\begin{abstract}
The density-matrix renormalization-group technique is used to calculate
the spin correlation functions $\langle S^x_jS^x_k\rangle$ and
$\langle S^z_jS^z_k\rangle$ of the one-dimensional $S=\frac{1}{2}$ XXZ
model in the gapless regime.
The numerical results for open chains of 200 spins are analyzed by
comparing them with correlation functions calculated from a low-energy
field theory.
This gives precise estimates of the amplitudes of the correlation
functions in the thermodynamic limit.
The exact amplitude recently conjectured by Lukyanov and Zamolodchikov
and the logarithmic correction in the Heisenberg model are confirmed
numerically.
\end{abstract}

\begin{multicols}{2}

The quantum spin chains have been a subject of very active research over
the many years.
Among others, the one-dimensional $S=\frac{1}{2}$ XXZ model,
\begin{equation}
{\cal H}=\sum^{L-1}_{j=1}
\left(
S^x_jS^x_{j+1}+S^y_jS^y_{j+1}+\Delta S^z_jS^z_{j+1}
\right),
\label{H}
\end{equation}
is a simplest nontrivial model and has served as a test ground of various
theoretical techniques.
These include the exact calculation via the Bethe ansatz,\cite{Baxter}
numerical computations such as the exact diagonalization
studies,\cite{Bonner} and the conformal field theory (CFT).\cite{review}
Although each approach alone is often not powerful enough to provide
sufficient information we need, employing these techniques together has
proven to be very successful to study, e.g., finite-size energy
spectra and spin correlation functions.
It is well known that the equal-time correlation functions of the XXZ
chain show the power-law decay in the critical regime ($-1<\Delta\le1$)
at zero temperature.
In the thermodynamic limit ($L\to\infty$) they have the asymptotic form
\begin{mathletters}
\begin{eqnarray}
\langle S^x_jS^x_k\rangle&=&
(-1)^{j-k}\frac{A_x}{|j-k|^\eta}
-\frac{\widetilde{A}_x}{|j-k|^{\eta+\frac{1}{\eta}}},
\label{SxSx}
\\
\langle S^z_jS^z_k\rangle&=&
(-1)^{j-k}\frac{A_z}{|j-k|^{1/\eta}}
-\frac{1}{4\pi^2\eta(j-k)^2},
\label{SzSz}
\end{eqnarray}
\end{mathletters}\noindent
where only the leading oscillating and non-oscillating terms are
written.  
The exact expression of the parameter $\eta$ appearing in the exponents,
\begin{equation}
\eta=1-\frac{1}{\pi}\cos^{-1}\Delta,
\label{eta}
\end{equation}
was obtained by comparing the result of the effective-field-theory
description (abelian bosonization) with the Bethe ansatz
solution.\cite{Luther} 
On the other hand, little had been known\cite{XY} on the correlation
amplitudes, until recently Lukyanov and Zamolodchikov\cite{LZ} (LZ)
conjectured the exact form of $A_x$:
\end{multicols}
\begin{equation}
A_x^{\rm LZ}=
\frac{(1+\xi)^2}{8}
\left[\frac{\Gamma(\frac{\xi}{2})}
           {2\sqrt{\pi}\,\Gamma(\frac{1}{2}+\frac{\xi}{2})}
\right]^\eta
\exp\left\{
-\int^\infty_0\frac{dt}{t}
 \left(\frac{\sinh(\eta t)}{\sinh(t)\cosh[(1-\eta)t]}
       -\eta e^{-2t}\right)\right\},
\label{eq:LZ}
\end{equation}
\begin{multicols}{2}\noindent
where $\xi=\eta/(1-\eta)$.
This result was then used in Refs.~\onlinecite{Affleck,Lukyanov} to
obtain the correlation amplitude for the antiferromagnetic
Heisenberg model ($\Delta=1$) with the logarithmic correction predicted
earlier from the renormalization-group
argument.\cite{Fink,AGSZ,Giamarchi,Singh}
The aim of this paper is to numerically determine the correlation
amplitudes ($A_x$, $\widetilde{A}_x$, and $A_z$) for broad ranges
of the anisotropy $\Delta$ in the critical regime.
To this end we apply the density-matrix renormalization-group (DMRG)
method\cite{White1,White2} to compute the spin correlation functions for
spin chains of $L=200$, and fit the numerical data to functional forms
expected from the effective field theory (abelian bosonization).
Indeed, the idea of fitting the correlation functions of finite periodic
systems to the CFT form has been applied successfully to the Heisenberg
case ($\Delta=1$) to examine the
logarithmic correction.\cite{Sandvik,Hallberg,Koma}
In the present study we use open spin chains instead of periodic ones,
not only because of the better performance of the DMRG method for open
chains,  but because the boundary effects provide useful information.
We numerically verify the LZ formula (\ref{eq:LZ}) and give numerical
data for the yet analytically unknown amplitudes $\widetilde{A}_x$ and
$A_z$.

We consider a finite XXZ chain of $L$ spins with open boundaries,
Eq.~(\ref{H}).
We assume $L$ to be an even integer so that the ground state is singlet.
According to the standard abelian bosonization
method,\cite{review,Eggert} 
the low-energy dynamics of the XXZ model is described by the Gaussian
model
\begin{equation}
H_0=\frac{1}{2}\int^{L+1}_0 dx\left[
\left(\frac{d\phi}{dx}\right)^2
+
\biggl(\frac{d\tilde\phi}{dx}\biggr)^2
\right],
\label{H_0}
\end{equation}
where $\phi(x)$ and $\tilde\phi(x)$ are bosonic fields.
We identify the site index $j$ with the continuous variable $x$, and
impose boundary conditions $\phi(0)=\phi(L+1)={\rm const.}$\ to respect
the fact that there is no spin at $j=0$, $L+1$.
The fields have the mode expansions
\begin{mathletters}
\begin{eqnarray}
\phi(x)&=&
\pi R+\frac{\widehat{Q}x}{L+1}
+\sum^\infty_{n=1}\frac{\sin(q_nx)}{\sqrt{\pi n}}
 \left(a_n+a_n^\dagger\right),
\label{phi}
\\
\tilde\phi(x)&=&
\tilde\phi_0
+i\sum^\infty_{n=1}\frac{\cos(q_nx)}{\sqrt{\pi n}}
 \left(a_n-a_n^\dagger\right),
\label{tilde-phi}
\end{eqnarray}
\end{mathletters}\noindent
where $q_n=\pi n/(L+1)$, $R=(\eta/2\pi)^{1/2}$,
$[\tilde\phi_0,\widehat{Q}]=i$, and
$[a_m,a_n^\dagger]=\delta_{m,n}$.
This leads to the commutation relation
$[\phi(x),\tilde\phi(y)]=-(i/2)[1+{\rm sgn}(x-y)]$.
The ground state $|0\rangle$ is the singlet vacuum state of
the bosons: $a_n|0\rangle=\widehat{Q}|0\rangle=0$.
The spin operators in the original Hamiltonian (\ref{H}) can be
expressed in terms of the phase fields as
\begin{mathletters}
\begin{eqnarray}
S^z_j&=&
\frac{1}{2\pi R}\frac{d\phi(x)}{dx}
+a(-1)^j\sin\frac{\phi(x)}{R},
\label{S^z}
\\
S^-_j&=&
e^{-2\pi iR\tilde\phi(x)}\left[
b\sin\frac{\phi(x)}{R}+c(-1)^j\right],
\label{S^-}
\end{eqnarray}
where $a$, $b$, and $c$ are real constants.
It follows that
\begin{equation}
S^x_j=
c(-1)^j\cos[2\pi R\tilde\phi(x)]
-ib\sin[2\pi R\tilde\phi(x)]\sin\frac{\phi(x)}{R}.
\label{S^x}
\end{equation}
\end{mathletters}\noindent
Note that the second term in Eq.\ (\ref{S^x}) is hermitian because
$[\phi(x),\tilde\phi(x)]=-i/2$.
These formulas are slightly modified from those in
Refs.~\onlinecite{Eggert,Ng,Affleck2} for reasons which will become
clear below. 
It is then straightforward to evaluate the correlation functions
for the ground state $|0\rangle$, yielding
\end{multicols}
\begin{mathletters}
\begin{eqnarray}
\langle S^x_jS^x_k\rangle&=&
\frac{f_{\eta/2}(2j)f_{\eta/2}(2k)}{f_\eta(j-k)f_\eta(j+k)}
\left\{
(-1)^{j-k}\frac{c^2}{2}
-\frac{b^2}{4f_{1/2\eta}(2j)f_{1/2\eta}(2k)}
 \left[\frac{f_{1/\eta}(j+k)}{f_{1/\eta}(j-k)}
       +\frac{f_{1/\eta}(j-k)}{f_{1/\eta}(j+k)}
 \right]
\right.\nonumber\\
&&\left.\hspace*{3.3cm}
-\frac{bc}{2}{\rm sgn}(j-k)\left[
 \frac{(-1)^j}{f_{1/2\eta}(2k)}-\frac{(-1)^k}{f_{1/2\eta}(2j)}
 \right]
\right\}
\label{SxSx-open}
\\
\langle S^z_jS^z_k\rangle&=&
\frac{(-1)^{j-k}a^2}{2f_{1/2\eta}(2j)f_{1/2\eta}(2k)}
\left[\frac{f_{1/\eta}(j+k)}{f_{1/\eta}(j-k)}
      -\frac{f_{1/\eta}(j-k)}{f_{1/\eta}(j+k)}\right]
-\frac{1}{4\pi^2\eta}
 \left(\frac{1}{f_2(j-k)}+\frac{1}{f_2(j+k)}\right)
\nonumber\\
&&
-\frac{a}{2\pi\eta}\left\{
\frac{(-1)^j}{f_{1/2\eta}(2j)}[g(j-k)+g(j+k)]
-\frac{(-1)^k}{f_{1/2\eta}(2k)}[g(j-k)-g(j+k)]
\right\}
\label{SzSz-open}
\end{eqnarray}
\end{mathletters}\noindent
\begin{multicols}{2}\noindent
with
\begin{mathletters}
\begin{eqnarray}
f_\alpha(x)&=&
\left[
\frac{2(L+1)}{\pi}\sin\left(\frac{\pi|x|}{2(L+1)}\right)
\right]^\alpha,
\label{f(x)}\\
g(x)&=&
\frac{\pi}{2(L+1)}\cot\left(\frac{\pi x}{2(L+1)}\right).
\label{g(x)}
\end{eqnarray}
\end{mathletters}\noindent
In deriving Eqs.\ (\ref{SxSx-open}) and (\ref{SzSz-open}) we used the
following regularization:
$\sum^\infty_{n=1}[1-\cos(q_nx)]/n=\ln[f_1(x)]$.
Note that Eqs.\ (\ref{SxSx-open}) and (\ref{SzSz-open}) reduce to
Eqs.\ (\ref{SxSx}) and (\ref{SzSz}) with $A_z=a^2/2$, $A_x=c^2/2$,
and $\widetilde{A}_x=b^2/4$ in the limit $L\to\infty$ with
$|j-\frac{L}{2}|\ll L$ and $|k-\frac{L}{2}|\ll L$.
It is also important to observe that Eqs.\ (\ref{SxSx-open}) and
(\ref{SzSz-open}) have exactly the same $j$- and $k$-dependence at the
Heisenberg point ($\eta=1$) in accordance with the SU(2) spin symmetry.
In fact, under the given boundary conditions, it is essential to use the
boson representation (\ref{S^z}) and (\ref{S^x}) to have this SU(2)
property.
Equation (\ref{SzSz-open}) has another nice feature that, when
$\eta=1/2$ and $a=1/\pi$, it coincides with the exact correlation
function $\langle S^z_jS^z_k\rangle$ of the XY model.
We use Eqs.\ (\ref{SxSx-open}) and (\ref{SzSz-open}) to analyze the
numerical data.

\begin{figure}
\epsfxsize=80mm
\epsfbox{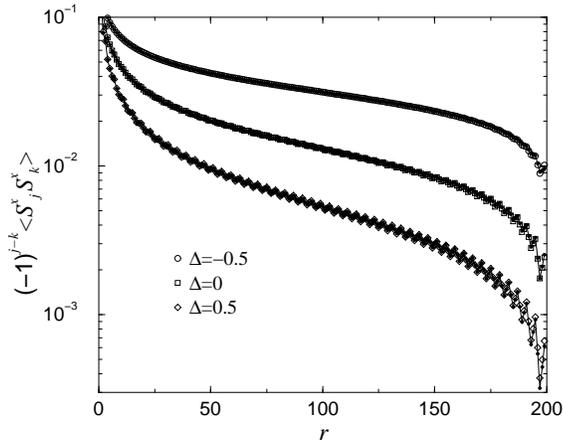}
\caption{\protect\narrowtext
$(-1)^{j-k}\langle S^x_jS^x_k\rangle$ versus $r=|j-k|$ for $\Delta=-0.5$,
0, and 0.5 (from above).
The open symbols are the DMRG data and the small dots connected by lines
are the fits, Eq.\ (\protect\ref{SxSx-open}).
}
\label{fig:cxx_fit}
\end{figure}

\begin{figure}
\epsfxsize=80mm
\epsfbox{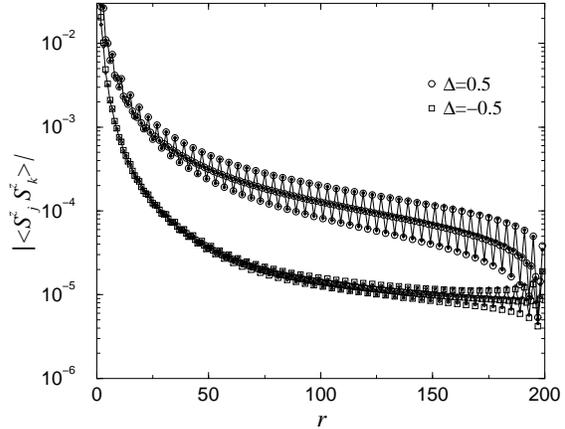}
\caption{\protect\narrowtext
$|\langle S^z_jS^z_k\rangle|$ versus $r=|j-k|$ for $\Delta=0.5$ (above)
and $-0.5$.
The open symbols are the DMRG data and the small dots connected by lines
are the fits, Eq.\ (\protect\ref{SzSz-open}).
}
\label{fig:czz_fit}
\end{figure}

We calculated the spin-spin correlation functions
$\langle S^\alpha_jS^\alpha_k\rangle$ ($\alpha=x,z$) for the ground
state of $L=200$ spin chains using the DMRG method with the improved
algorithm.\cite{White1,White2} 
We employed the finite system method, and the maximum number of kept
states $m$ is 80.
From the difference between the data at $m=80$ and $m=60$,
we estimate the numerical error due to the truncation of the Hilbert space
to be of order $10^{-5}$ for $\langle S^x_jS^x_k\rangle$ and $10^{-6}$ for
$\langle S^z_jS^z_k\rangle$.
Figures \ref{fig:cxx_fit} and \ref{fig:czz_fit} show some of our
numerical data of the correlations between $S^\alpha_j$ and $S^\alpha_k$,
where $j=(L-r+1)/2$ and $k=(L+r+1)/2$ for odd $r$ and $j=(L-r)/2$ and
$k=(L+r)/2$ for even $r$.
The DMRG results are shown by open symbols, whose size is, however, made
larger than the numerical error for illustration.
Taking $a$, $b$, and $c$ as free parameters and using the exact exponent
Eq.\ (\ref{eta}),  we fit the data\cite{fitting} using
Eq.~(\ref{SxSx-open}) and (\ref{SzSz-open}).
The small dots connected by lines in Figs.~\ref{fig:cxx_fit} and
\ref{fig:czz_fit} are the results of the fitting.
From the relation $A_x=c^2/2$, $\widetilde{A}_x=b^2/4$, and
$A_z=a^2/2$, we obtain the correlation amplitudes, and the results are
summarized in Table I.
The excellent agreement between the LZ's conjecture $A^{\rm LZ}_z$ and the
value obtained from the fitting clearly demonstrates that Eq.\
(\ref{eq:LZ}) is indeed exact, and that there is little finite-size
correction left in $a$, $b$, and $c$ because Eqs.\ (\ref{SxSx-open}) and
(\ref{SzSz-open}) correctly account for most of the finite-size
effect.
We also note that the numerical estimate of $A_z$ at $\Delta=0$ agrees
with the exact value, $1/2\pi^2$.
For $\Delta\lesssim-0.7$, it is difficult to estimate $\widetilde{A}_x$
and $A_z$ because the exponent $1/\eta$ for these terms becomes so large
that they give only a negligible contribution to the correlation
functions.
As seen in Figs.~\ref{fig:cxx_fit} and \ref{fig:czz_fit},
for $-1<\Delta\lesssim0.6$ Eqs.\ (\ref{SxSx-open}) and (\ref{SzSz-open})
can fit the numerical data extremely well including the oscillations
of period 4, which are characteristic of open chains.

\begin{figure}
\epsfxsize=80mm
\epsfbox{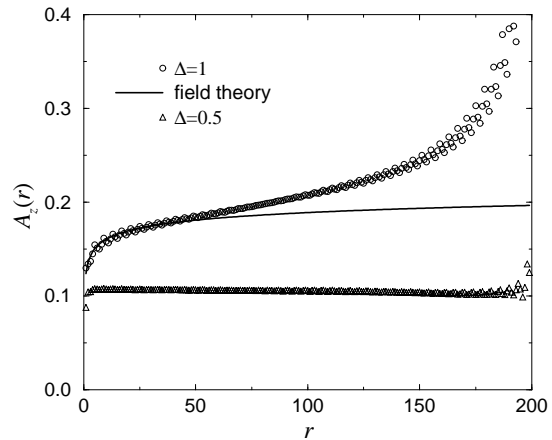}
\caption{\protect\narrowtext
Effective correlation amplitude $A_z(r)$ as a function of $r$ for
$\Delta=1$ and 0.5.
The solid line represents the $\protect\sqrt{\log r}$ correction
[Eq.\ (\protect\ref{A_z-2})] predicted from a field-theory calculation.
$A_z(r)$ is almost constant ($\sim0.1$) at $\Delta=0.5$.
}
\label{fig:Az}
\end{figure}

As $\Delta$ becomes closer to 1, however, the quality of the fitting
becomes poor.
This is due to the irrelevant operator $\cos(2\phi/R)$ ignored in Eq.\
(\ref{H_0}).
Roughly speaking, this operator gives rise to corrections of order
$r^{-4(1/\eta-1)}$ to the amplitudes, and at $\Delta=1$ this
turns into the logarithmic correction.\cite{Affleck,Lukyanov}
Since this type of correction is not taken into account in
Eqs.\ (\ref{SxSx-open}) and (\ref{SzSz-open}), they cannot describe the
correlation functions completely for the parameter regime where the
effect of the leading irrelevant operator becomes important.
To elucidate this, we plot in Fig.~\ref{fig:Az} the effective
amplitude $A_z(r)$ given by $A_z(r)=a^2(r)/2$, where $a(r)$ is
obtained by solving Eq.\ (\ref{SzSz-open}) for each $r=|j-k|$.
We see that $A_z(r)$ is constant for $\Delta=0.5$, which is another
indication that the fitting works well.
At $\Delta=1$, on the other hand, $A_z(r)$ is roughly an increasing
function of $r$.
This result is compared with the recent analytical result valid in the
limit $L\to\infty$,\cite{Lukyanov}

\begin{equation}
A_z(r)=\frac{1}{\sqrt{8\pi^3g}}
\left(1-\frac{3}{16}g^2+\frac{156\zeta(3)-73}{384}g^3\right),
\label{A_z-2}
\end{equation}
where $g$ is determined from
\begin{equation}
\frac{1}{g}+\frac{1}{2}\log g
=\log\left(2\sqrt{2\pi}\,e^{\gamma+1}r\right)
\end{equation}
with $\gamma=0.5772\ldots$.
Equation (\ref{A_z-2}) is shown as a solid line in Fig.~\ref{fig:Az},
which goes on top of the numerical data for $r\lesssim50$.
Thus, our numerical result is consistent with the recent analytic
calculation,\cite{Affleck,Lukyanov}
$\langle S^\alpha_jS^\alpha_k\rangle=
[\log(2\sqrt{2\pi}e^{\gamma+1}r)/8\pi^3]^{1/2}/r$ for $r\gg1$.
The discrepancy at larger $r$ in Fig.~\ref{fig:Az} would be due to the
finite-size effect or the boundary effect.
The numerical estimates of the amplitudes for $\Delta=0.8$ and 0.9 also
suffer from the correction coming from the irrelevant operator and have
larger error bars than for $\Delta\lesssim0.6$.

In summary, we have demonstrated that the spin correlation functions of
open XXZ chains in the critical regime can be fitted very well with the
correlation functions calculated from the effective low-energy theory
(free bosons), if the effect of the irrelevant operator is not
severe.
From the fitting, we obtained the correlation amplitudes for broad
ranges of the parameter $\Delta$ with high precision and numerically
verified the Lukyanov and Zamolodchikov's conjecture.
We hope that our numerical estimates for $A_z$ and $\widetilde{A}_x$
will be confirmed in turn by analytic calculations in the future.
The numerical computation was carried out at the
Yukawa Institute Computer Facility, Kyoto University.
A.F.\ is supported by a Monbusho Grant for overseas research.

\end{multicols}

\widetext

\begin{table}
\caption{Correlation amplitudes estimated from the fitting of the
numerical data.
$A^{\rm LZ}_x$ is the conjectured amplitude, Eq.~(\protect\ref{eq:LZ}).
The figures in parentheses indicate the error bar on the last quoted
digits.
}
\label{tbl}
\begin{tabular}{ddddd}
$\Delta$ & $A_x^{\rm LZ}$ & $A_x$ & $\widetilde{A}_x$ & $A_z$ \\
\hline
$-$0.9 & 0.15567 & 0.15560(3) & ***      & *** \\
$-$0.8 & 0.15904 & 0.1589(1) & ***       & *** \\
$-$0.7 & 0.15968 & 0.1595(1) & ***       & 0.008(1) \\
$-$0.6 & 0.15912 & 0.1589(2) & 0.0048(14) & 0.0133(1) \\
$-$0.5 & 0.15786 & 0.1576(2) & 0.0076(9)  & 0.0184(4) \\
$-$0.4 & 0.15617 & 0.1558(2) & 0.0099(7)  & 0.0235(2) \\
$-$0.3 & 0.15417 & 0.1538(3) & 0.0122(4)  & 0.02921(3) \\
$-$0.2 & 0.15196 & 0.1515(3) & 0.0144(2)  & 0.03556(3) \\
$-$0.1 & 0.14958 & 0.1491(3) & 0.0164(2)  & 0.0425(2) \\
   0.0 & 0.14709 & 0.1466(4) & 0.0182(2)  & 0.0501(5) \\
   0.1 & 0.14451 & 0.1440(4) & 0.01995(7) & 0.0588(3) \\
   0.2 & 0.14187 & 0.1414(3) & 0.02154(5) & 0.0683(6) \\
   0.3 & 0.13921 & 0.1388(3) & 0.02296(4) & 0.0791(8) \\
   0.4 & 0.13656 & 0.1362(3) & 0.02420(4) & 0.0918(9) \\
   0.5 & 0.13400 & 0.1339(1) & 0.02525(6) & 0.1063(9) \\
   0.6 & 0.13164 & 0.1320(1) & 0.0261(2)  & 0.1236(5) \\
   0.7 & 0.12973 & 0.1308(5) & 0.0267(3)  & 0.145(1) \\
   0.8 & 0.12896 & 0.132(1)  & 0.0271(7)  & 0.171(5) \\
   0.9 & 0.13214 & 0.138(3)  & 0.027(2)   & 0.20(1) \\
\end{tabular}
\end{table}


\begin{references}
\bibitem{fitting} For the least-squares fit, we have typically used the
data of $20\le r\le160$.
The error arising from the fitting procedure is estimated by changing the
region of the data used for the fitting.
\end{references}
\end{document}